\documentclass[12pt]{article}
\usepackage{cite}
\usepackage{graphicx}
\usepackage{amssymb,amsfonts}

\renewcommand{\theequation}
{\arabic{section}.\arabic{equation}}

\makeatletter
\def\eqnarray{ \stepcounter{equation} \let\@currentlabel=\theequation
 \global\@eqnswtrue
 \global\@eqcnt\z@
 \tabskip\@centering
 \let\\=\@eqncr
 $$\halign to \displaywidth\bgroup\@eqnsel\hskip\@centering
 $\displaystyle\tabskip\z@{##}$&\global\@eqcnt\@ne
 \hfil$\displaystyle{{}##{}}$\hfil
 &\global\@eqcnt\tw@$\displaystyle\tabskip\z@{##}$\hfil
 \tabskip\@centering&\llap{##}\tabskip\z@\cr}
\makeatother

\makeatletter
\def\@arrayacol{\edef\@preamble{\@preamble \hskip .5\arraycolsep}}
\def\array{\let\@acol\@arrayacol \let\@classz\@arrayclassz
\let\@classiv\@arrayclassiv \let\\\@arraycr\def\@halignto{}\@tabarray}
\makeatother
\makeatletter
\newcounter{subeqncnt}
\def\thesubeqncnt{\alph{subeqncnt}}
\def\subequations{\begingroup%
   \stepcounter{equation}\edef\@tempa{\theequation}%
   \let\c@equation\c@subeqncnt\c@subeqncnt\z@
   \edef\theequation{\@tempa\noexpand\thesubeqncnt}}

\makeatother
\newcommand{\captionfonts}{\small}
\makeatletter % Allow the use of @ in command names
\long\def\@makecaption#1#2{%
\vskip\abovecaptionskip
\sbox\@tempboxa{{\captionfonts #1: #2}}%
\ifdim \wd\@tempboxa >\hsize {\captionfonts #1: #2\par} \else
\hbox to\hsize{\hfil\box\@tempboxa\hfil}%
\fi \vskip\belowcaptionskip}
\makeatother % Cancel the effect of \makeatletter
\newcommand{\del}{\partial}

\newcommand{\dd}{{\rm d}}

\def\be{\begin{equation}}
\def\ee{\end{equation}}
\newcommand{\bea}{\begin{eqnarray}}
\newcommand{\eea}{\end{eqnarray}}

\begin{document}

\setlength{\baselineskip}{7mm}

\begin{flushright}
{\tt SHU-Pre2020-04} \\
\end{flushright}

\vspace{1cm}

\begin{center}
{\Large Causality  of black holes in 4-dimensional Einstein-Gauss-Bonnet-Maxwell theory}

\vspace{1cm}

{\sc{Xian-Hui Ge}}$^{*\flat}$, {\sc{Sang-Jin Sin}}$^{\dagger }$

$*${\it{Department of Physics, Shanghai University},} \\
{\it{Shanghai 200444, China}} \\
{\sf{gexh@shu.edu.cn}}
\\
$\dagger$ {\it{Department of Physics,}} {\it{Hanyang University,}}
{\it{Seoul 133-791, Korea}} \\
{\sf{sjsin@hanyang.ac.kr}}
\\
$\flat$ {\it{Center for Gravitation and Cosmology, College of Physical Science and Technology,
 }} {\it{Yangzhou University,}}
{\it{Yangzhou 225009, China}}
\end{center}

\vspace{1.5cm}

\begin{abstract}
We study charged black hole solutions in 4-dimensional (4D) Einstein-Gauss-Bonnet-Maxwell theory to the linearized perturbation level. We first compute the shear viscosity to entropy density ratio.
We then demonstrate  how bulk causal structure analysis imposes a upper bound on the Gauss-Bonnet coupling constant in the AdS space. Causality  constrains the value of Gauss-Bonnet coupling constant $\alpha_{GB}$ to be bounded by $\alpha_{GB}\leq 0$ as $D\rightarrow 4$.
\end{abstract}

\vspace{1cm}

%%%%%
\section{Introduction}

 The AdS/CFT
correspondence \cite{ads/cft,gkp,w} provides a powerful tool  for studying the physics of strongly coupled gauge theories and also can be used to examining alternative theories to the general relativity. Alternative theories to Einstein's General Relativity   paradigm can be scrutinized by diverse approaches. The higher derivative gravity with $\alpha'$ corrections
was also studied widely within the framework of the AdS/CFT correspondence (see \cite{kp,shenker,shenker1,cai1,gmsst,gs,cai2,cai3,buchel} for an incomplete list). For Einstein-Gauss-Bonnet (EGB) theory, strong constraints can be imposed on the Gauss-Bonnet coupling constant from the analysis of the bulk causal structure. For 5-dimensional (5D) EGB theory, causality demands $\alpha_{GB}\leq 0.09$ to avoid superluminal propagation of signals in the dual boundary field theory \cite{shenker,shenker1,buchel,gs}. Recently, revived interests on EGB gravity in 4-dimensional spacetime have been first concerned in \cite{lin20}.

% The result of RHIC experiment on the viscosity/entropy ratio turns out to be in favor of the prediction of AdS/CFT ~\cite{pss0,kss,bl}. Some attempt has been made to map the entire process of RHIC experiment in terms of
%gravity dual \cite{ssz}.

%The way to include chemical potential in the theory was figured out in~\cite{ksz,ht}. Phases of these theories were also discussed in \cite{nssy1,kmmmt,nssy2,huang,ht}.

The 4-dimensional EGB gravity is realized by first rescaling the coupling constant $\alpha'\rightarrow \frac{\alpha'}{D-4}$ of the Gauss-Bonnet term and then take the limit $D\rightarrow 4$ \cite{lin20}. In this way, the Lovelock's theorem \cite{lovelock,lovelock1,lancos} can be bypassed and spherically symmetric 4D black hole solutions can be obtained in the presence of the Gauss-Bonnet term.

In this paper, we are going to investigate whether causality violation happens in the $4D$ EGB gravity and check the upper bound of the Gauss-Bonnet coupling constant. As shown in \cite{shenker,shenker1,gmsst,gs}, higher derivative terms in the gravity action can result in superluminal propagation of gravitons outside the light cone of a given background geometry. The graviton cone in such case does not coincide with the standard null cone or light cone defined by the background metric. Utilizing the tool provided by AdS/CFT correspondence, firstly we will study the linearized perturbation of the black holes in 4D Einstein-Gauss-Bonnet-Maxwell (EGBM) gravity. Then we calculate the shear viscosity to entropy density ratio in this context.  We then examine the causality constraint  on the Gauss-Bonnet coupling constant.
 % For 5D black holes in EGB theory, the coupling constant is constrained by $\alpha_{GB} \leq 0.09$ \cite{shenker,shenker1}.
 We are going to show that for 4D black holes in EGBM theory, if we take the limit $D\rightarrow 4$  before the series expansion of the local speed of the transverse graviton on the boundary, the speed cannot reach the local speed of light (i.e. $c=1$) on the boundary unless we choose $\alpha_{GB}=0$. The local speed of graviton is smaller than the local speed of light for any positive value of  $\alpha_{GB}$.  Meanwhile, if the limit $D\rightarrow 4$ is taken after the series expansion of the local speed of the transverse graviton near the boundary, no causality violation requires $\alpha_{GB} \leq 0$.

 We will show that the bulk graviton propagates  faster than the local speed of light could leads to signals in the boundary theory propagate outside the light cone. According to the AdS/CFT correspondence, the boundary theory is non-gravitational. In a boosted frame, perturbations will propagate backward in time. Hence, these could lead to unambiguous signals of causality violation.

 The structure of this paper is organized as follows. In section 2, we  study the charged black hole solutions in 4D  Einstein-Gauss-Bonnet-Maxwell theory.  Then, in section 3, we compute the shear viscosity to entropy density ratio. In section 4, we discuss the bulk causal structure and its boundary consequences.  The conclusion and discussions are provided in the last section.

% We also explain how the geodesic motion of the graviton in the bulk to be interpreted as the group velocity on the boundary.

\section{Charged black hole solutions in  Einstein-Gauss-Bonnet-Maxwell theory}
We now consider the Einstein-Maxwell Gauss-Bonnet in D dimensions with a negative cosmological constant $\Lambda_0=-\frac{(D-1)(D-2)}{l^2}$ given by the action
\begin{equation}
\label{action} I=\frac{1}{16 \pi}\!\int\!\dd^{D}\!x
\sqrt{-g}\Big(R-2\Lambda_0+\frac{\alpha'}{D-4} \mathcal{G}-
F_{\mu\nu}F^{\mu\nu}\Big),
\end{equation}
where $\alpha'$ is a (positive) Gauss-Bonnet coupling constant with
dimension $\rm(length)^2$, the field strength is defied as
$F_{\mu\nu}(x)=\del_\mu A_\nu(x)-\del_\nu A_\mu (x)$ and $\mathcal{G}=\left(R_{\mu\nu\rho\sigma}
R^{\mu\nu\rho\sigma}-4R_{\mu\nu}R^{\mu\nu}+R^2\right)$.
The general D-dimensional static and maximally symmetric black hole can be described as
\begin{equation}\label{metric2}
ds^2=-f(r)dt^2+\frac{dr^2}{f(r)}+r^2 d\Omega^2_{2,\kappa},~~~d\Omega^2_{2,\kappa}=\frac{dx^2}{1-\kappa x^2}+x^2 d\varphi^2,
\end{equation} and an electrostatic vector potential
\begin{equation}
A_t=V(r)dt,
\end{equation}
where $\kappa=-1,0,1$. Since all the functions are radially dependent only, by substituting  (\ref{metric2}) into the action, we obtain
\begin{equation}
V'(r)=-\frac{Q}{r^{D-2}},
\end{equation}
with the integral constant $Q$ as the electric charge. The metric function $f(r)$ can be obtained by defining a new variable $\psi(r)$
\be
f(r) \equiv \kappa-r^2 \psi(r).
\ee
In this form, the action reduces to
\be
I=\frac{\Omega_{d-2}}{16 \pi}\int dt dr (D-2)\bigg[r^{D-1}\psi(1+\alpha'(D-3)\psi)+\frac{r^{D-1}}{l^2}+\frac{2 Q^2 r^{3-D}}{(D-3)(D-2)}\bigg]',
\ee
with $\Omega_{d-2}=\frac{2 \pi^{\frac{D-2}{2}}}{\Gamma(\frac{D-1}{2})}$. Notice that $ \psi(r)$ satisfies the relation \cite{caiGB}
\be
\psi+\alpha'(D-3)\psi^2=\frac{16\pi M}{(D-2)r^{D-1}\Omega_{d-2}}-\frac{1}{l^2}-\frac{2 Q^2 r^{4-2D}}{(D-3)(D-2)},
\ee
where $M$ is the ADM mass.  We then obtain the metric function and the scalar potential as follows
\bea
f(r)&=&\kappa-\frac{r^2}{2\alpha' (D-3)}\bigg [-1\pm \sqrt{1-4 (D-3)\alpha'\bigg(\frac{1}{l^2}+\frac{2 Q^2 r^{4-2D}}{(D-3)(D-2)}-\frac{16\pi M r^{1-D}}{(D-2)\Omega_{D-2}}\bigg)}~\bigg ],\nonumber\\
A_t &=&-\frac{Q}{(D-3)r^{D-3}}dt.
\eea
The sign $+$ denotes perturbative branch in $\alpha'$, while the $-$ sign corresponds to the branch that the metric function $f(r)$ goes to infinity as $\alpha'\rightarrow 0$. We choose the $+$ sign hereafter.  A rigorous method of compactifying EGB gravity on a $(D-4)$-dimensional maximally symmetric space was introduced in \cite{lu20}. The thermodynamics and geometric properties of the 4D Einstein-Gauss-Bonnet black holes have been studied in several papers
\cite{konoplya20,Fernandes20,more20,Liu,Casalino,Kumar,Hegde,Doneva,Ghosh,Olea,Mahapatra,Fernandes,Singh:2020xju,Singh:2020nwo,Almendra,samart}.
In what follows, we focus on the black hole solution with the planar horizon by taking $\kappa=0$.

For planar black branes in AdS space, the line elements can be written as
\be \label{metric}
ds^2=-H(r)N^2 dt^2+H^{-1}(r) dr^2+\frac{r^2}{l^2}dx^i dx^i, ~~{\rm with }~~i=1,...,D-2,
\ee
where
\bea
H(r)&=&\frac{r^2}{2\alpha_{GB} l^2}\left[1-\sqrt{1-4\alpha_{GB}\bigg(1-\frac{ml^2}{r^{D-1}}+\frac{q^2l^2}{r^{2D-4}}\bigg)}~\right]\nonumber\\
&=&\frac{r^2}{2\alpha_{GB} l^2}\left[1-\sqrt{1-4\alpha_{GB}\bigg(1-\frac{r^{D-1}_{+}}{r^{D-1}}
-a\frac{r^{D-1}_{+}}{r^{D-1}}+a\frac{r^{2D-4}_{+}}{r^{2D-4}}\bigg)}~\right].
\eea
Note that  $\alpha_{GB}$ and $\alpha'$ are connected by a relation
$\alpha_{GB}=(D-3)\alpha'/l^2$,
$a=\frac{q^2l^2}{r^{2D-4}_{+}}$ denotes dimensionless charge parameter and the parameter $l$ corresponds to
AdS radius. The horizon is located at $r=r_{+}$. The gravitational
mass $M$ and the charge $Q$ are expressed as
\begin{eqnarray*}
M&=&\frac{(D-2) \Omega_{D-2}}{16 \pi   }m,
\\
Q^2&=&\frac{ (D-2)(D-3)}{ 2 }q^2.
\end{eqnarray*}
Taken the limit $\alpha'\rightarrow 0$, the solution recovers the metric of Reissner-Nordstr\"om-AdS black branes.

The constant $N^2$ in the metric (\ref{metric}) can be determined at the
boundary whose geometry would reduce to the flat Minkowski metric
conformaly, i.e.\ $\dd s^2\propto -c^2\dd t^2+\dd\vec{x}^2$. On the
boundary with $r\rightarrow\infty$, we have
$$
H(r)N^2 \rightarrow\frac{r^2}{l^2},
$$
so that $N^2$ is fixed as
\begin{equation}
N^2=\frac{1}{2}\Big(1+\sqrt{1-4 \alpha_{GB}}\ \Big).\label{N}
\end{equation}
Note that the boundary speed of light is specified to be unity
$c=1$.

The Hawking temperature at the event horizon is given by
\begin{equation}
T=\frac{1}{2\pi\sqrt{g_{rr}}}\frac{\dd \sqrt{g_{tt}}}{\dd
r}=\frac{Nr_+}{4\pi l^2}\left[(D-1)-(D-3)a\right].
\end{equation}
 The black brane approaches
extremal as $a\rightarrow \frac{D-1}{D-3}$ (i.e.\ $T\rightarrow
0$). The entropy density is given by \cite{caiGB}
\begin{equation}
s=\frac{1}{4 }\frac{r^{D-2}_{+}}{l^{D-2}}.
\end{equation}
In order to investigate  the causality structure and the upper bound of the Gauss-Bonnet coupling constant in four-dimensional spacetime, we will take the  $D \rightarrow 4$ limit and analysis the shear viscosity to entropy density ratio first.

\section{Shear viscosity}
In this section, we are going to study the shear viscosity in the 4D Einstein-Maxwell Gauss-Bonnet gravity theory and examine the shear viscosity bound. Since we already took $\alpha'\rightarrow \frac{\alpha'}{D-4}$ in equation (\ref{action}),  we will compute the shear viscosity in  general $D$ dimensions and then take the limit $D\rightarrow 4$ so as to  circumvent the Lovelock theorem.
It is convenient to introduce new coordinates in the following
computation
\begin{eqnarray}
&&z=\frac{r}{r_{+}},
~~\omega=\frac{l^2}{r_{+}}\bar{\omega},~~k_{3}=\frac{l^2}{r^2_{+}}\bar{k}_3,~~
f(z)=\frac{l^2}{r^2_{+}}H(r), \nonumber\\
&&f(z)=\frac{z^2}{2
\alpha_{GB}}\bigg[1-\sqrt{1-4\alpha_{GB}\bigg(1-\frac{a+1}{z^{D-1}}+\frac{a}{z^{2D-4}}}\bigg)\bigg].
\end{eqnarray}
We now study the tensor type perturbation
$h^{x_i}_{x_j}(t,x_i,z)=\phi(t,x_i,z)$ with $ i\neq j$ on the black brane background of
the form
$$
ds^2=-f(z)N^2\dd t^2+\frac{\dd z^2}{b^2 f(z)}+\frac{z^2}{b^2
l^2}\left(2\phi(t,x_i,z)\dd x_i \dd x_j+\sum^{D-2}_{i=1}\dd
x^2_{i}\right),
$$
where $b=\frac{1}{r^2_{+}}$.
 Using Fourier decomposition
$$
\phi(t, x_i,z) = \!\int\!\frac{\dd^{D-1}k}{(2\pi)^{D-1}}
\mbox{e}^{-i\bar{\omega} t+i\bar{k}_{i}x_i}\phi(k, z),
$$
we can obtain the equation of motion for
$\phi(z)$ from the Einstein-Gauss-Bonnet-Maxwell field equation
\begin{equation}\label{maineq}
\partial_z\bigg(\mathcal{N}^{~zx_i}\partial_z \phi\bigg)+\omega^2 \mathcal{N}^{~tx_i}\phi-k^2_i \mathcal{N}^{~x_i x_j}\phi=0,
\end{equation}
where
\begin{eqnarray}
&&\mathcal{ N}^{zx_i}=\frac{1}{16 \pi}\sqrt{-g}g^{x_i x_i} g(z), \nonumber\\
&& \mathcal{ N}^{tx_i}=-\frac{1}{16 \pi}\sqrt{-g}g^{tt} g(z),\nonumber\\
&& \mathcal{ N}^{x_i x_j}=\frac{1}{16 \pi}\sqrt{-g}g^{x_i x_i}g_2(z),\nonumber\\
&&g(z)=1-\frac{2 \alpha_{GB}}{D-3}
\left[z^{-1}f'+z^{-2}(D-5)f\right],\nonumber
\\
&&g_2(z)=1-\frac{2\alpha_{GB}}{(D-3)(D-4)}\left(f''+(D-5)(D-6)z^{-2}f+2(D-5)z^{-1}f'\right),
\end{eqnarray}
and the prime denotes the derivative with respect to $z$. Note that the factors  $(D-5)$ and $(D-6)$ in the expression of $g_2(z)$ comes from higher than 5-dimensional contribution of the Gauss-Bonnet theory.

The Green function related to the shear viscosity takes the form
\be
G_{x_i x_j,x_i x_j}=\frac{\mathcal{N}^{zx_i}\partial_z \phi}{\phi}.
\ee
The shear viscosity can be defined as
\be
\eta_{x_i x_j,x_i x_j}=\frac{-G_{x_i x_j,x_i x_j}}{i \omega}.
\ee
We can then recast equation (\ref{maineq})  as a flow equation
\be
\partial_z \eta_{x_i x_j,x_i x_j}=\bigg(\frac{\eta^2_{x_i x_j,x_i x_j}}{\mathcal{N}^{zx_i}}-\mathcal{N}^{t x_i}\bigg)+\frac{i}{\omega} \mathcal{N}^{x_ix_j} k^2_i.
\ee
The shear viscosity can be computed by requiring horizon regularity
\be
\eta_{x_i x_j,x_i x_j}=\bigg(\mathcal{N}^{zx_i}\mathcal{N}^{tx_i}\bigg)\bigg|_{z=1}=\frac{1}{16 \pi
}\left(\frac{r^{D-2}_{+}}{l^{D-2}}\right)\Big(1-\frac{2\alpha_{GB}}{(D-3)}
[(D-1)-(D-3)a]\Big).
\ee
The ratio of the shear viscosity to the entropy density for 4D charged black hole solutions in Gauss-Bonnet gravity is then
\begin{equation}
\frac{\eta_{x_ix_j,x_ix_j}}{s} =\frac{1}{4 \pi } \left(1-\frac{2\alpha_{GB}}{(D-3)}
[(D-1)-(D-3)a]\right).
\end{equation}
In the limit $D\rightarrow 4$, we obtain
\be
\frac{\eta_{x_ix_j,x_ix_j}}{s} =\frac{1}{4 \pi } \left[1-{2\alpha_{GB}}
(3-a)\right].
\ee
We can see that for 4D Einstein-Gauss-Bonnet theory, the shear viscosity bound can still be violated. But as the black hole temperature approaches zero $a\rightarrow 3$, one can recover the well-known result $\eta/s \sim 1/4 \pi$ \cite{kovtun,PKS,kss,bl,ssz,ksz}.
\section{Bulk causal structure }
 According to the AdS/CFT correspondence, the physics in bulk 4D AdS gravity is dual to boundary 3D quantum field theory on its boundary. In this section, we study the bulk causal structure and show how a high-momentum metastable state in the bulk graviton wave equation that may have consequence for boundary causality.

 Because of higher derivative terms in the gravity action, the equation
(\ref{maineq}) for the propagation of a transverse graviton differs
from the standard Klein-Gordon equation of a minimally coupled massless scalar field propagating
in the same background geometry. Writing the wave function of the transverse graviton as
\begin{equation}
\label{phi} \phi(x_i,z)=\mbox{e}^{-i\omega t+ikz+ik_{i}x_{i}},
\end{equation}
and taking large momenta limit $k^\mu\rightarrow\infty$, one can
find that the equation of motion (\ref{maineq}) reduces to
\begin{equation}
\label{effeq} k^{\mu}k^{\nu}g_{\mu\nu}^{\rm eff}\simeq 0,
\end{equation}
where the effective metric is
\begin{equation}
\dd s^2_{\rm eff} =g^{\rm eff}_{\mu\nu}\dd x^\mu\dd x^\nu ={N^2
f(r)} \left(-\dd t^2+\frac{1}{c^2_g}\dd x^2_i\right)
+\frac{1}{f(r)}\dd r^2.
\end{equation}
Note that the function $c_g$ can be interpreted as the local speed of graviton on a constant $r$-hypersurface \cite{gs,gs10}:
\begin{equation}
c^2_g(z)=\frac{N^2
f}{z^2}\frac{1-\frac{2\alpha_{GB}}{(D-3)(D-4)}\left(f''+(D-5)(D-6)z^{-2}f+2(D-5)z^{-1}f'\right)}{1-\frac{2
\alpha_{GB}}{(D-3)} \left[z^{-1}f'+z^{-2}(D-5)f\right]}.
\end{equation}
The local speed of light defined by the background metric $c^2_b=\frac{N^2f(z)}{z^2}$, which is $1$ at the boundary $z\rightarrow \infty$. In the bulk, the background local speed of light $c_b$ is smaller than $1$ because of the redshift of the black hole geometry.

The causality problem arises because a graviton wave packet  moving at speed $c_g$ in the bulk  corresponds to perturbations of the stress tensor propagating with the same velocity in the boundary theory.
Since the replacement $\alpha'\rightarrow \frac{\alpha'}{D-4}$ was done already, now we can expand the local speed of graviton  $c^2_g$ near the
boundary $z \rightarrow\infty$ in the limit $D\rightarrow 4$,
\be\label{cgone}
c^2_g = \frac{1}{\sqrt{1-4 \alpha_{GB}}}+\left[-\frac{1+\sqrt{1-4\alpha_{GB}}-12\alpha_{GB}}{2(1-4\alpha_{GB})^{3/2}}+\frac{12 \alpha_{GB}}{(D-4)(1-4\alpha_{GB})}\right]\frac{1+a}{z^3}+\mathcal{O}(z^{-4}).
\ee
The first term in (\ref{cgone}) is not equal to the speed of light.  This is because higher dimension term with factors $(D-5)$ and $(D-6)$ also contribute to $c^2_g$ at the leading order when expand near the boundary and modify the leading order term.
The second term in the square brackets diverges as $D\rightarrow 4$, which implies that the limiting procedure does not work well to the linearized perturbation level \cite{lu20}.
As the local speed of graviton should not be bigger than $1$ (the local
speed of light of the boundary CFT), the leading term in (\ref{cgone}) should satisfy
\be
\frac{1}{\sqrt{1-4\alpha_{GB}}}\leq 1.
\ee
It can be satisfied by requiring $\alpha_{GB}\geq 0$.  However, if one requires the local speed of graviton to be exactly $1$, one should take $\alpha_{GB}=0$.
%
%\begin{equation}
%-\frac{(D^2-5D+10)}{2(D-3)}+\frac{(D-1)}{(D-3)(1-4\alpha)}
%-\frac{1}{2\sqrt{1-4\alpha}}\leq 0.
%\end{equation}
%

An alternative approach to  the limit $D\rightarrow 4$ is  to expand $c^2_g$ near the boundary $z \rightarrow \infty$ for general $D$ \cite{gs}
\begin{eqnarray} \label{cg429}
c^2_g-1=&&
\left(-\frac{(D^2-5D+10)(1+a)}{2(D-3)(D-4)}\right.\nonumber\\&&
\left.+\frac{(D-1)(1+a)}{(D-3)(D-4)(1-4\alpha_{GB})}
-\frac{1+a}{2\sqrt{1-4\alpha_{GB}}}\right)\frac{1}{z^{D-1}}
+\mathcal{O}(z^{-D}).
\end{eqnarray}
The condition that the local speed of graviton should be smaller than $1$  requires
\begin{equation}
-\frac{(D^2-5D+10)}{2(D-3)(D-4)}+\frac{(D-1)}{(D-3)(D-4)(1-4\alpha_{GB})}
-\frac{1}{2\sqrt{1-4\alpha_{GB}}}\leq 0.
\end{equation}
Note that this is significantly different from the result obtain in \cite{gs}
\begin{equation}\label{old}
\alpha_{GB} \leq
\frac{D^4-10D^3+41D^2-92D+96}{4(D^2-5D+10)^2},
\end{equation}
 which in the $D\rightarrow 4$ limit leads to $\alpha_{GB} \leq 0$. But as $D=5$, one can recover the well-known result $\alpha_{GB}\leq 0.09$ \cite{shenker1}.
 %There is also one important difference between the new constraint shown in (\ref{cons}) and the one (\ref{old}) without the $\frac{\alpha'}{D-4}$ limit: Once the replacement   $\alpha'\rightarrow \frac{\alpha'}{D-4}$ is done in the action, equation (\ref{cons}) does not agree with  equation (\ref{old}) for $D\geq 6$. But as $D=5$, both of them reduce to $\alpha\leq 0.09$.

 The bulk causal structure and its relation with the boundary theory can be discussed as follows. In the boundary theory, the local operators create bulk disturbances at infinity that propagate along the graviton geodesics deep inside the bulk. The equation of motion for $\phi$ in  (\ref{maineq}) can be interpreted as an equation describing metastable quasiparticles of the boundary field theory.   Now, we recast the equation of motion of the wave function  (\ref{maineq})  in a Schr$\rm \ddot{o}$dinger form,
\begin{equation}
-\frac{d^2 \psi}{dr^2_{*}}+V\left(z(r_{*})\right)\psi=\omega^2 \psi,
~~~\frac{dr_{*}}{dz}=\frac{1}{Nf(z)},\label{schr}
\end{equation}
where $\psi\left(z(r_{*})\right)$ and the potential is defined by
\begin{eqnarray}
&&\psi =K(z)\phi,~~~K(z)\equiv\sqrt{\frac{g(z)}{z^{D-2}f(z)}},
V=k^2c^2_g+V_{1}(z),\nonumber\\ &&V_{1}(z)\equiv
N^2\left[\left(f(z)\frac{\partial \ln K(z)}{\partial
z}\right)^2+f(z)\frac{\partial}{\partial z}\left(f(z)\frac{\partial
\ln K(z)}{\partial z}\right)\right].
\end{eqnarray}
Geodesics starting from the boundary can bounce back to the boundary. It has been proven that the quasiparticles can travel faster than the speed of light and violate causality \cite{gs}.
 From the geodesic  equation of motion
\begin{equation}
g^{\rm eff}_{\mu\nu}\frac{\dd x^{\mu}}{\dd s}\frac{\dd x^{\nu}}{\dd
s}=0,
\end{equation}
and the Bohr-Sommerfield quantization condition
\begin{equation}
 \int \dd r_{\ast}\sqrt{\omega^2-k^2 c^2_g}=(n-\frac{1}{4})\pi,
\end{equation}
one can find that the group velocity of the test particle along the
geodesic line is given by \cite{shenker1}
\begin{equation}
 v_g=\frac{\dd \omega}{\dd k}\rightarrow c_g.
\end{equation}
For 4D EGBM theory, the equation  (\ref{cgone}) shows that $c_g$ is slower than the speed of light for $\alpha_{GB}\geq 0$, so do the group velocity of the test particle.  However, equation (\ref{old}) implies that causality impose the condition
$\alpha_{GB}\leq 0$ to avoid propagation of signals faster than the speed of light. A natural question is whether  $\alpha_{GB}$ can be negative from the holographic point of view. For black holes in 5D EGB gravity,  $\alpha_{GB}$ has a lower bound $\alpha_{GB}\geq -7/36$ from the analysis of sound mode perturbations \cite{buchel}.  We leave the study on sound mode perturbations in 4D EGB gravity theory to a future work.

%Now we can conclude that as dimensions of
%space-time go up, causality restricts the value of $\lambda$ in the
%region $\lambda\leq 1/4$.  In next section, we will prove that in
%the extremal limit $a\rightarrow \frac{D-1}{D-3}$, the stability of
%the black brane also requires that  $\lambda$ should also be bounded
%by $1/4$ in the limit $D\rightarrow \infty$.

\section{Conclusions and discussions}
\setcounter{equation}{0} \setcounter{footnote}{0} In summary, we studied the linearized metric perturbation of black holes in 4D Einstein-Gauss-Bonnet-Maxwell theory within the framework of the AdS/CFT correspondence. The charged black hole solutions were obtained for general $D$ dimensions. We then study the shear viscosity to entropy density ratio by considering the planar black brane solution.  The ratio ${\eta_{x_ix_j,x_ix_j}}/{s} =\frac{1}{4 \pi } \left[1-{2\alpha_{GB}} (3-a)\right]$ turns out to be different from the Kovtun-Son-Starinets bound $\eta/s=1/4\pi$ if $\alpha_{GB}$ is non-vanishing.  We then investigated the bulk causal structure of the 4D charged black holes.  In order to guarantee no violation of causality in the boundary field, the Gauss-Bonnet coupling $\alpha_{GB}$ should be in the range $\alpha_{GB}\leq 0$. Note that these results were obtained by following the procedure proposed in \cite{lin20}: First define the Gauss-Bonnet coupling $\alpha'\rightarrow \frac{\alpha'}{D-4}$ and then take $D\rightarrow 4$ limit.

There are some subtleties in the bulk causal structure analysis.  In (\ref{cgone}), we expanded the local speed of graviton at boundary $z\rightarrow \infty$ by setting $D=4$. But the leading term does not match with the local speed of light and there is a divergent term in (\ref{cgone}) at $\mathcal{O}(z^{-3})$ order.  A way to bypass this situation is to expand $c_g$ near
the boundary for general $D$ dimensions as given in (\ref{cg429}). In this case, causality requires $\alpha_{GB}\leq 0$, which is a strong constraint on the reasonable value of $\alpha_{GB}$.  It would be  useful  to  consider a mathematically more rigorous definition for the $D\rightarrow 4$ limit of EGB gravity at the linearized level.

If one adds a linear axion  field into the action (\ref{action}) and break the translational symmetry, then the bulk causal structure could be drastically changed. For example, in 5D EGB theory,  causality violation still  happens in the presence of the linear scalar field  but with an  effective mass of the graviton dependence \cite{hartnoll,wang16,Sadeghi}.  If the effective mass of the graviton is large enough, then there will be no causality violation and hence no constraints for the Gauss-Bonnet coupling. For 4D EGB gravity with a linear axion field, one may expect the same result.  We defer these discussions to a future study.

\vspace*{10mm} \noindent
 {\large{\bf Acknowledgments}}

\vspace{1mm} We would like to thank Chunshan Lin and Qingbing Wang for helpful discussions. This work is partly supported by NSFC (No.11875184). SJS is supported by Mid-career Researcher Program through the National Research Foundation of Korea grant No. NRF-2016R1A2B3007687.


\begin{thebibliography}{99}

\bibitem{ads/cft}
J. M. Maldacena, {Adv. Theor. Math. Phys.} {\bf 2} (1998) 231, {\tt
[arXiv:hep-th/9711200]}.
\bibitem{gkp}
S. S. Gubser, I.R. Klebanov and A.M. Polyakov, Phys.\ Lett.\ {\bf
B428} (1998) 105, {\tt [arXiv:hep-th/9802109]}.
\bibitem{w}
 E. Witten, Adv.\ Theor.\ Math.\ Phys.\ {\bf 2} (1998) 253, {\tt
[arXiv:hep-th/9802150]}.

\bibitem{kp}
Y. Kats and P. Petrov, JHEP 0901 (2009) 044 {\tt
[arXiv:0712.0743[hep-th]]}.
\bibitem{shenker}
M. Brigante, H. Liu, R.C. Myers, S. Shenker and S. Yaida, Phys. Rev.
{\bf D77} (2008) 126006, {\tt [arXiv:0712.0805[hep-th]]}.
\bibitem{shenker1}
M. Brigante, H. Liu, R.C. Myers, S. Shenker and S. Yaida, Phys. Rev.
Lett. {\bf 100} (2008) 191601, {\tt [arXiv:0802.3318[hep-th]]}.

\bibitem{gs}
X. H. Ge and S. J. Sin, Shear viscosity, %instability and the upper bound of the Gauss-Bonnet coupling constant,
 JHEP 05 (2009) 051 [arXiv:0903.2527]
\bibitem{cai1} R. G. Cai and Y. W. Sun, JHEP {\bf 0603} (2008) 052, {\tt [arXiv:0807.2377[hep-th]]}

\bibitem{gmsst}
X.~H. Ge, Y. Matsuo, F.-W. Shu, S.-J. Sin and T. Tsukioka, JHEP 0810
(2008) 009, {\tt [arXiv:0808.2354[hep-th]]}

\bibitem{cai2}R. G. Cai, Z. Y. Nie and Y. W. Sun, Phys. Rev. {\bf D78} (2008)126007  {\tt [arXiv:0811.1665[hep-th]]}
\bibitem{cai3}R. G. Cai, N. Ohta, Z. Y. Nie and Y. W. Sun, {\tt [arXiv:0901.1421[hep-th]]}
\bibitem{buchel}A. Buchel and R.C. Myers, %Causality of Holographic Hydrodynamics,
JHEP 08 (2009) 016 [arXiv:0906.2922]
\bibitem{lin20}
D. Glavan and C. Lin, %¡°Einstein-Gauss-Bonnet gravity in 4-dimensional space-time,¡±
Phys. Rev. Lett., vol. 124, no. 8, p. 081301, 2020.
\bibitem{lovelock}
D. Lovelock, %¡°The Einstein tensor and its generalizations,¡±
J. Math. Phys. 12 (1971) 498.
\bibitem{lovelock1} D. Lovelock, %¡°The four-dimensionality of space and the einstein tensor,¡±
J. Math. Phys. 13 (1972) 874.
\bibitem{lancos} C. Lanczos, % ¡°A Remarkable property of the Riemann Christoffel tensor in four dimensions,¡±
 Annals Math. 39
(1938) 842.
\bibitem{caiGB}
R.G. Cai, Phys. Rev. {\bf D65} (2002) 084014, {\tt
[arXiv:hep-th/0109133]};
\\
R.G. Cai and Q. Guo, Phys. Rev. {\bf D69} (2004) 104025, {\tt
[arXiv:hep-th/0311020]}.\\
R.G. Cai, Phys. Lett. B {\bf 582} (2004) 237, {\tt
[arXiv:hep-th/0311240]}.
\bibitem{lu20} H. L$\rm \ddot{u}$ and Y. Pang, %¡°Horndeski gravity as $D\rightarrow 4$ limit of Gauss-Bonnet,¡±
arXiv: 2003.11552[hep-th].
\bibitem{konoplya20}
R. A. Konoplya and A. Zhidenko, %Black holes in the four-dimensional Einstein-Lovelock gravity,
{\tt
[arXiv:2003.07788[gr-qc]]}.
\bibitem{Fernandes20}
P. G. S. Fernandes,% Charged Black Holes in AdS Spaces in 4D Einstein Gauss-Bonnet Gravity,
{\tt
[arXiv:2003.05491[gr-qc]]}.
\bibitem{more20}
R. Konoplya and A. Zinhailo, arXiv:2003.01188 [gr-qc]; M. Guo and P.C. Li,
arXiv:2003.02523 [gr-qc].
\bibitem{Liu}
S.W. Wei and Y.X. Liu, arXiv:2003.07769
[gr-qc].
\bibitem{Casalino} A. Casalino, A. Colleaux, M. Rinaldi and S. Vicentini, arXiv:2003.07068 [gr-qc].
\bibitem{Kumar} R. Kumar and S.G. Ghosh, arXiv:2003.08927 [gr-qc].
\bibitem{Hegde} K. Hegde, A.N. Kumara,
C.L.A. Rizwan, A.K.M. and M.S. Ali, arXiv:2003.08778 [gr-qc].
\bibitem{Doneva} D. D. Doneva and
S.S. Yazadjiev, arXiv:2003.10284 [gr-qc].
\bibitem{Ghosh} S. G. Ghosh and S.D. Maharaj, arXiv:2003.09841
[gr-qc].
\bibitem{Olea} R. Araneda, R. Aros, O. Miskovic and R. Olea, %¡°Magnetic mass in 4D AdS gravity,¡±
arXiv:1602.07975

\bibitem{Mahapatra} S. Mahapatra, %¡°A note on the total action of 4D Gauss-Bonnet theory,¡±
arXiv: 2004.09214.

\bibitem{Fernandes}P. G. Fernandes, P. Carrilho, T. Clifton and D. J. Mulryne, %¡°Derivation of Regularized Field Equations for the Einstein-Gauss-Bonnet Theory in Four Dimensions,¡±
[arXiv:2004.08362 [gr-qc]].
\bibitem{Singh:2020xju}
  D.~V.~Singh and S.~Siwach,
  %``Thermodynamics and P-v criticality of Bardeen-AdS Black Hole in 4-D Einstein-Gauss-Bonnet Gravity,''
  arXiv:2003.11754 [gr-qc].

\bibitem{Singh:2020nwo}
D.~V.~Singh, S.~G.~Ghosh and S.~D.~Maharaj,
%``Clouds of string in the novel $4D$ Einstein-Gauss-Bonnet black holes,''
[arXiv:2003.14136 [gr-qc]].

\bibitem{Almendra} A. Arag$\rm \acute{o}$n, R. B$\rm \acute{e}$car, P. A. Gonz$\rm\acute{a}$lez, Y. V$\rm\acute{a}$squez,
%``Perturbative and nonperturbative quasinormal modes of 4D Einstein-Gauss-Bonnet black holes,''
arXiv:2004.05632 [gr-qc].

    \bibitem{samart} D. Samart and  P. Channuie,  arXiv: 2005.02826[gr-qc].

\bibitem{kovtun}
P. Kovtun, D.T. Son and A.O. Starinets, Phys.\ Rev.\ Lett.\  {\bf
94} (2005) 111601, {\tt [arXiv:hep-th/0405231]}.

\bibitem{PKS}
G. Policastro, D. T. Son and A.O. Starinets, Phys.\ Rev.\ Lett.\
{\bf 87} (2001) 081601, {\tt [arXiv:hep-th/0104066]}.
\bibitem{kss}
P. Kovtun, D. T. Son and A.O. Starinets,
JHEP {\bf 0310} (2003) 064, \\
{\tt [arXiv:hep-th/0309213]}.

\bibitem{bl}
A. Buchel and J. T. Liu,
Phys.\ Rev.\ Lett.\  {\bf 93} (2004) 090602, \\
{\tt [arXiv:hep-th/0311175]}.
\bibitem{ssz}
E. Shuryak, S.-J. Sin and I. Zahed,
J.\ Korean Phys.\ Soc.\  {\bf 50} (2007) 384, \\
{\tt [arXiv:hep-th/0511199]}.
\bibitem{ksz}
K.-Y. Kim, S.-J. Sin and I. Zahed, {\tt [arXiv:hep-th/0608046]}.
\bibitem{gs10}
 X. H. Ge, S. Sin, S. Wu, and G. Yang, % Shear viscosity and instability from third order Lovelock gravity,
  Phys. Rev. D
80, 104019 (2009).
\bibitem{hartnoll}
S. A. Hartnoll, D. M. Ramirez, and J. E. Santos, % Entropy production, viscosity bounds and bumpy black holes,
J. High Energy Phys. 03 (2016) 170.
\bibitem{wang16}
Y. Wang and X. H. Ge,
Phys. Rev. {\bf D 94} (2016)  066007.
\bibitem{Sadeghi}
M. Sadeghi and S. Parvizi, %¡°Hydrodynamics of a black brane in Gauss-Bonnet massive gravity,¡± 
Class. Quant. Grav. 33 (2016)  035005 
[arXiv:1507.07183 [hep-th]].


%\bibitem{ht}
%N. Horigome and Y. Tanii, JHEP {\bf 0701} (2007) 072, {\tt
%[arXiv:hep-th/0608198]}.
%\bibitem{nssy1}
%S. Nakamura, Y. Seo, S.-J. Sin and K.P. Yogendran, {\tt
%[arXiv:hep-th/0611021]}.


%\bibitem{kp}
%Y. Kats and P. Petrov, JHEP 0901 (2009) 044 {\tt
%[arXiv:0712.0743[hep-th]]}.
%\bibitem{shenker}
%M. Brigante, H. Liu, R.C. Myers, S. Shenker and S. Yaida, Phys. Rev.
%{\bf D77} (2008) 126006, {\tt [arXiv:0712.0805[hep-th]]}.
%\bibitem{shenker1}
%M. Brigante, H. Liu, R.C. Myers, S. Shenker and S. Yaida, Phys. Rev.
%Lett. {\bf 100} (2008) 191601, {\tt [arXiv:0802.3318[hep-th]]}.
%\bibitem{cai1} R. G. Cai and Y. W. Sun, JHEP {\bf 0603} (2008) 052, {\tt [arXiv:0807.2377[hep-th]]}
%\bibitem{neupane}
%I.P. Neupane and N. Dahhich, {\tt [arXiv:0808.1919[hep-th]]}.
%\bibitem{gmsst}
%X.~H. Ge, Y. Matsuo, F.-W. Shu, S.-J. Sin and T. Tsukioka, JHEP 0810
%(2008) 009, {\tt [arXiv:0808.2354[hep-th]]}
%\bibitem{cai2}R. G. Cai, Z. Y. Nie and Y. W. Sun, Phys. Rev. {\bf D78} (2008)126007  {\tt [arXiv:0811.1665[hep-th]]}
%\bibitem{cai3}R. G. Cai, N. Ohta, Z. Y. Nie and Y. W. Sun, Phys.\ Rev.\ {\bf D79} (2009) 066004 {\tt [arXiv:0901.1421[hep-th]]}
%\bibitem{dotti}
%G. Dotti and R.J. Gleiser, Phys.\ Rev.\ {\bf D72} (2005) 044018,
%{\tt [arXiv:gr-qc/0503117]}; \\
%R.J. Gleiser and G. Dotti, Phys.\ Rev.\  {\bf D72} (2005) 124002,
%{\tt [arXiv:gr-qc/0510069]}; \\
%M. Beroiz, G. Dotti and  R.J. Gleiser,
%Phys. Rev. {\bf D76} (2007) 024012, \\
%{\tt [arXiv:hep-th/0703074]}.
%\bibitem{konoplya} R.A. Konoplya and A. Zhidenko, Phys. Rev. {\bf D77} (2008) 104004, {\tt [arXiv:0802.0267]}.



\end{thebibliography}
\end{document}